\documentclass[prb,aps,showpacs,floatfix,twocolumn]{revtex4}
\usepackage{times}
\usepackage{amsmath,bm,amsfonts}
\usepackage{graphicx}
\usepackage{color}

\begin{document}

\title{Oxygen-vacancy-induced charge carrier in n-type interface of
    LaAlO$_{3}$ overlayer on SrTiO$_{3}$ (001): interface vs bulk doping carrier}

\author{Yun Li$^{1}$}
\author{S. Na Phattalung$^{2}$}
\author{S. Limpijumnong$^{2}$}
\author{Jaejun Yu$^{1}$}
\email[Corresponding author.\ ]{Email: jyu@snu.ac.kr}
\address{$^{1}$Department of Physics and Astronomy, FPRD, Center for Strongly Correlated Materials Research,
Seoul National University, Seoul 151-747, Korea \\
$^{2}$School of Physics, Suranaree University of Technology, Nakhon Ratchasima 30000, Thailand\\
}

\date{\today}

\begin{abstract}
We investigated the role of oxygen vacancy in $n$-type interface
of LaAlO$_3$ (LAO) overlayer on SrTiO$_3$ (STO) (001) by carrying
out density-functional-theory calculations. Comparing the total
energies of the configurations with one vacancy in varying
locations we found that oxygen vacancies favor to appear first in
LAO surface. These oxygen vacancies in the surface generate a
two-dimensional distribution of carriers at the interface,
resulting in  band bending at the interface in STO side. Dependent
on the concentration of oxygen vacancies in LAO surface, the
induced carrier charge at the interface partially or completely
compensates the polar electric field in LAO. Moreover, the
electronic properties of oxygen vacancies in STO are also
presented. Every oxygen vacancy in STO generates two electron
carriers, but this carrier charge has no effect on screening polar
field in LAO. Band structures at the interface dependent on the
concentrations of oxygen vacancies are presented and compared with
experimental results.
\end{abstract}


\maketitle

\section{Introduction}

Since Ohtomo and Hwang reported the existence of a high mobility
electron gas at the $n$-type (LaO)$^{+}$/(TiO$_{2}$)$^{0}$
interface between two band-gap insulators LAO and
STO\cite{ohtomo04}, many novel properties related to the electron
gas at the interface, such as insulator-metal transition,
superconductivity, and ferromagnetism were
found\cite{thiel06,cen08,rijnders08,reyren07,Caviglia08,brinkman07}.
The mechanism of conductivity and the dimensionality of the
induced carrier at the interface were intensively studied. In the
$n$-type interface structure a polar electrical field along LAO
[001] direction is arisen from the alternating stack of
(LaO)$^{+}$ and (AlO$_{2}$)$^{-}$ charged layers. Charge
reconstruction at the interface was proposed as a way to avoid
diverging electrostatic potential with thick LAO, which
compensates the polar electric field in LAO and induces electron
gas at the interface\cite{ohtomo04,nakaga06}. The densities and
distributions of the induced carriers charge  were measured in
various experiments
\cite{ohtomo04,thiel06,brinkman07,basletic08,nakaga06,eckstein07,
herranz07,siemons07,kalabukhov07,willmott07,yoshimatsu08,sing09}.
Two different origins of the induced carriers charge, i.e.
intrinsic and extrinsic doping, have been proposed. Intrinsic
charge doping at the interface mainly happens in stoichiometrical
structure without oxygen
vacancy\cite{thiel06,brinkman07,basletic08, willmott07,sing09}.
With thick LAO overlayer the charge transfers from valence band of
LAO surface to conduction band of STO at the interface, leading to
a metallic interface\cite{ pentcheva09,son09, li09}. While
extrinsic charge doping was found to be ascribed to oxygen
vacancies in many experiments\cite{ohtomo04,brinkman07,
herranz07,kalabukhov07,Maurice08}.

Noticeably,  the conducting properties at intrinsic and extrinsic
doped interfaces measured in the experiments are very different.
In the ideal interface of thin LAO overlayer without oxygen
vacancy the carrier densities measured by Hall effect and x-ray
photoelectron spectroscopy (XPS) are even less than 0.1 electron
per two-dimensional unit cell
(2-d.u.c.)\cite{thiel06,basletic08,sing09}. Recent
density-functional-theory (DFT) calculations showed that ionic
polar distortions  partially compensate polar electric field in
LAO, reducing carrier density at the
interface\cite{pentcheva06,pentcheva08,pentcheva09,ishibashi,lee08,li09}.
For thin LAO overlayer the carrier charge was found in XPS
experiment to distribute in a few layers of STO at the
interface\cite{sing09}. Latest DFT study found that for less than
7 layers of LAO the carrier charge distributes in 3 layers of STO
but for thicker LAO the carrier could accumulate in deep STO
layers about 3 nm away from the interface. However, with oxygen
vacancies in the interface structure  the carrier densities
measured in experiments vary from 0.5 to three orders of magnitude
e/2-d.u.c. dependent on the oxygen pressure in preparing
process\cite{ohtomo04,brinkman07,herranz07,kalabukhov07}. The
experiment in Ref.[\onlinecite{herranz07}] explicitly verified
that high carrier density is ascribed to high concentration of
oxygen vacancies in STO substrate which were generated under lower
oxygen pressure during preparing the sample.


Besides the carrier density, another issue which people have been
debating extensively is the distribution of carrier charge induced
by oxygen vacancy. In some sample produced in high oxygen pressure
two-dimensional  distributions were
observed\cite{thiel06,brinkman07,basletic08,sing09}, but in others
produced in low oxygen pressure obvious 3-dimensional distribution
in STO were observed \cite{basletic08,herranz07}. These
experimental hints indicated the carrier distribution is related
to the concentration of oxygen vacancies in some way. Up to now,
people have realized that oxygen vacancy play an important role in
n-type LAO/STO interface, however the locations of oxygen
vacancies and the dependence of two-dimensional and
three-dimensional carrier distributions on the locations of oxygen
vacancies are still not clear.


In this paper we studied the n-type interface with oxygen
vacancies by using density-functional-theory (DFT) calculations.
It should be noted that the interface composed of sandwich
structure was also investigated in experiment but the structure of
LAO overlayer on STO substrate has attracted the most of
attentions because of its abundant conducting properties. In this
paper we focused on the structure of LAO overlayer on STO. We
calculated and compared the total energies of the configurations
with varying locations and concentrations of oxygen vacancies.  We
found that the locations of oxygen vacancies are related to the
concentration. From the calculations of electronic structures we
also unveiled that the distribution of carriers is strongly
dependent on the locations of oxygen vacancies.  Based on the
results from our calculations, energy band structures at the
interface and localized surface conductivity were discussed.

\begin{figure}[htbp]
    \centering
    \includegraphics[width=0.3\textwidth]{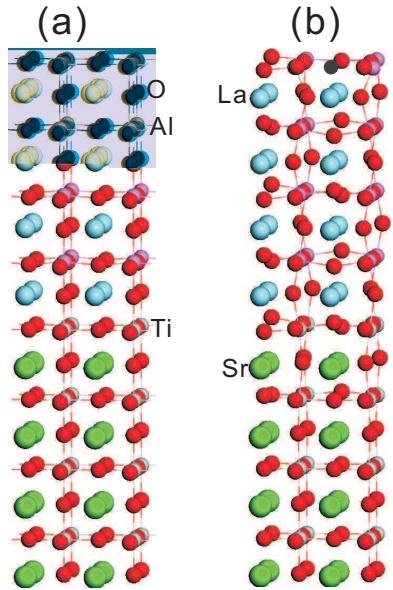}
    \caption{(Color online) Atomic structures of (2$\times$2) supercells
     (a) without oxygen vacancy, and (b) with oxygen vacancy in LAO surface.
      Oxygen vacancy is denoted by black ball.}\label{Struc}
\end{figure}

\section{Computational method}

We carried out DFT calculations by using the Vienna \textit{ab
initio} Simulation Package (VASP) \cite{kresse96} within a
generalized gradient approximation \cite{wang91} together with the
projector augmented wave pseudopotentials \cite{blochl94,kresse96}
and the cut-off energy of 400 eV for the plane wave basis.  We
modeled the LAO/STO interface by a slab consisting of 2 to 7  LAO
layers on top of  STO(001) substrate and a vacuum region of 14
{\AA} along the $c$-axis in a supercell geometry. Dipole
corrections were used to correct the errors of electrostatic
potential, forces, and total energy caused by periodic boundary
condition \cite{makov95}. To simulate various concentrations of
oxygen vacancies the in-plane unit-cells were taken as
(2$\times$2), (3$\times$2), and (3$\times$3). $\Gamma$-centered
(5$\times$5), (3$\times$5), and (3$\times$3) \textbf{k}-point
meshes were used to sample the Bullion zone, respectively.
Fig.~\ref{Struc} (a) and (b) present relaxed (2$\times$2)
supercells  without oxygen vacancy and with oxygen vacancy in LAO
surface, respectively. The in-plane lattice constant of the slab
was constrained at the calculated equilibrium lattice constant
$a=3.942$ {\AA} of the STO substrate. All coordinates of atomic
positions were fully relaxed with forces less than 0.02eV/{\AA}
except for the atoms in the bottom two layers of STO, which were
fixed in their bulk positions.

\section{Location of oxygen vacancies of low concentration}
In this paper we divide the concentrations of oxygen vacancies in
the interface structure into two regions, i.e. low concentration
and high concentration. In the former the concentrations of oxygen
vacancies is less than or equal to 1/4 per 2-d.u.c. and the
density of induced carriers is less than or equal to 0.5 electron
per 2-d.u.c.. While in the latter the concentrations of oxygen
vacancies is larger than 1/4 vacancy per 2-d.u.c. and the density
of induced carriers is larger than 0.5 electron per 2-d.u.c.. To
simulate low concentration of oxygen vacancy in the interface
structure only one vacancy is involved in the supercells. While
for higher concentration more vacancies are involved in the
supercells. For convenient we use n$_{V}$ and n$_{c}$ to denote
the concentration of oxygen vacancies and carrier density in
2-d.u.c., respectively.

\subsection{Binding energies of oxygen atom}  

\begin{figure}[htbp]
    \centering
    \includegraphics[width=0.45\textwidth]{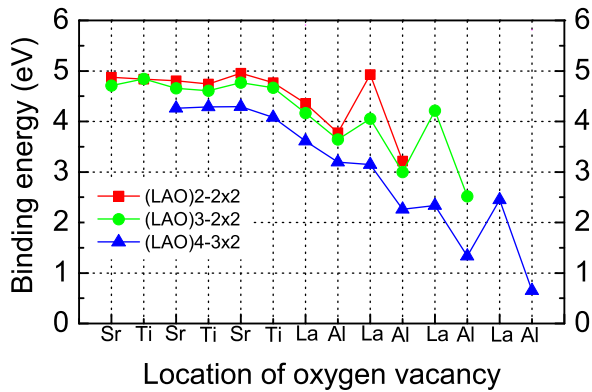}
    \caption{(Color online) Binding energies of oxygen
        atom corresponding to one vacancy at varying locations in (3$\times$2) supercell
        consisting of STO substrate and  4 LAO layers and
        (2$\times$2) supercells consisting of STO substrate and 2 or 3 LAO layers.}
        \label{BindingE1}
\end{figure}

Oxygen vacancies were explicitly confirmed to exist in STO
substrate of the interface structure produced under low oxygen
pressure\cite{basletic08,herranz07}. However, up to now  there has
been no report about whether oxygen vacancy exits in LAO. To
investigate the location of oxygen vacancy we calculated the total
energies of oxygen vacancy in varying locations under various
concentrations. We used (3$\times$2) and (2$\times$2) supercells
with one vacancy as the representatives of n$_{V}$ less than and
equal to 1/4 per 2-d.u.c., respectively. To compare them in same
plot we calculated the binding energy of one oxygen atom
corresponding to the vacancy defined as:
$$E_{b}=(E_{V}+1/2E_{O_{2}})-E_{0} , $$ in which $E_{V}$, $E_{O_{2}}$,
and $E_{0}$ are total energies of the system with oxygen vacancy,
oxygen molecule and ideal system without vacancy, respectively.
Figure~\ref{BindingE1} presents the binding energies of  one
oxygen atom corresponding to the vacancy at varying locations in
(3$\times$2) and (2$\times$2) supercell. Oxygen atom in LAO
surface has the lowest binding energy, which is far less than that
in STO substrate. This indicates that oxygen vacancy is most
easily formed in LAO surface rather than in STO substrate.
Moreover the binding energy of oxygen atom in LAO surface
decreases with the thickness of LAO. Such results are strongly
related to polar electric field in LAO. In the ideal interface
structure the polar  field shifts the energy bands of LAO toward
higher energy layer by layer, substantially raising the total
energy. While oxygen vacancy in LAO  dopes  electron carriers at
the interface, which partially or completely screen the polar
field in LAO, remarkably lowering the total energy. In contrast to
the vacancy in the surface, the vacancy inside LAO just screens
part of polar field from the layer with the vacancy to the
interface. Therefore corresponding configuration has higher total
energy, as illustrated in Fig.~\ref{BindingE1}.

\subsection{Formation energy of oxygen vacancy}
\begin{figure}[htbp]
    \centering
    \includegraphics[width=0.45\textwidth]{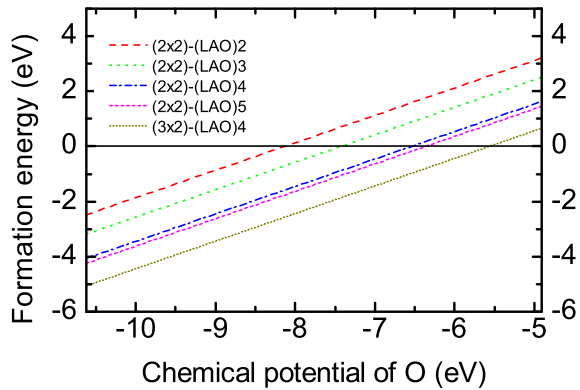}
    \caption{(Color online) Formation energies of one oxygen vacancy
     in LAO surface as a function of oxygen chemical potential.
     The thicknesses of LAO vary from 2 to 5 layers for
     (2$\times$2) supercells, and only 4 layers of LAO for
     (3$\times$2) supercell. Formation energy of ideal structure
      without vacancy is taken as zero.}\label{FormationE}
\end{figure}
To investigate dependence of the formation of oxygen vacancy in
the surface on the thickness of LAO, we calculated  the formation
energies of one vacancy in the surface of (2$\times$2) supercells
consisting of 2 to 5 layers of LAO according to the formula
$$E_{f}=E_{V}-(E_{0}-\mu_{O}) ,$$ in which  $\mu_{O}$ is chemical potential of oxygen
atom, $E_{V}$, and $E_{0}$ are total energies of the system with
one oxygen vacancy in the surface and ideal system without
vacancy, respectively. As shown in  Fig.~\ref{FormationE}
 the formation energies of (2$\times$2) supercells with one vacancy in LAO
surface decrease with the thickness of LAO overlayer and trend to
converge for those of thickness of LAO more than 4 layers. As
aforementioned, in the ideal structure the electrostatic energy
induced by polar field increases with the thickness of LAO.
Following electronic calculations show that for (2$\times$2)
supercells with one oxygen vacancy in the surface the density of
doping charge at the interface is 0.5 e/2-d.u.c., which exactly
compensates polar electric field.  This implies that oxygen
vacancy in the surface of thicker LAO overlayer eliminates more
electrostatic energy than that of thinner LAO overlayer.  While
once the LAO overlayer exceeds 4 layers, as demonstrated in
previous DFT studies, in ideal structure without vacancy the
intrinsic doping charge at the interface partially screens the
polar field, preventing the increasing of electrostatic energy
with the thickness of LAO\cite{pentcheva09,son09, li09}. Therefor,
the energy difference between vacant and ideal structures does not
increase with thickness of LAO. So we can predict that under same
preparing condition oxygen vacancy can be formed more easily in
the sample with thicker LAO film, while this tendency does not
strengthen any more once LAO exceeds 4 layers. Moreover, as shown
in Fig.~\ref{FormationE}, for two unitcells (2$\times$2) and
(3$\times$2) which have the same 4 layers of LAO the lower
concentration of vacancy is more stable and therefor more easily
be formed. As well known, oxygen pressure is a dominant factor of
the formation of oxygen vacancy in LAO/STO interface structure. In
Fig.~\ref{FormationE} at oxygen rich limit the ideal structure has
the lowest formation energy. This is consistent with the
experimental results that vacancy-free structures were fabricated
under high oxygen
pressure\cite{thiel06,brinkman07,basletic08,sing09}. While in the
middle or near the poor limit vacant structures have lower
formation energies.

\section{Electronic structures of oxygen vacancies of low concentration}
\subsection{Oxygen-vacancy state in LAO surface and induced carrier}
\begin{figure}[htbp]
    \centering
    \includegraphics[width=0.45\textwidth]{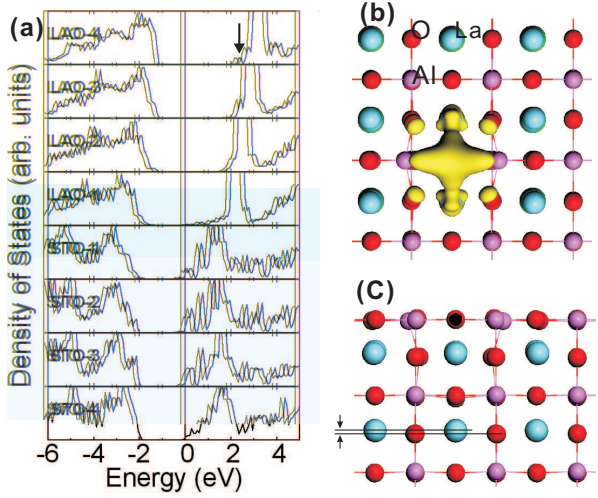}
    \caption{(Color online) (a) Layer-projected DOS for the (3$\times$3)
        supercell with one oxygen vacancy in LAO surface. (b) Spacial
        distribution of oxygen-vacancy state in the (3$\times$3) surface.
        (c) Side view of polar distortions of cations and anions in upper
        3 layers of LAO.} \label{figDOS:1}
\end{figure}

Above binding-energy calculations have shown that in the
configurations of n$_{V}<=$1/4 per 2-d.u.c. oxygen vacancy lies in
LAO surface. To investigate the electronic properties of oxygen
vacancy in the surface we calculated (2$\times$2), (3$\times$2),
(3$\times$3), and (4$\times$4) supercells with one vacancy,
respectively. As a representative of n$_{V}<$1/4 per 2-d.u.c., the
electronic structure and local atomic structure of (3$\times$3)
supercell are presented in Fig.~\ref{figDOS:1}. One can see in
Fig.~\ref{figDOS:1} (a) and (b) that the empty state of oxygen
vacancy lies a little bit below the conduction band minimum (CBM)
of LAO and is localized in the surface layer. Obviously, the level
of oxygen vacancy is higher than CBM of STO. Charge transfer from
the state of oxygen vacancy in the surface to conduction band of
STO at the interface lowers the total energy of the system. Our
calculation showed that 2 electrons accumulate at the interface,
i.e. n$_{c}$= 0.22 e/2-d.u.c., which is the twice of n$_{V}$.  The
same relation between n$_{c}$ and n$_{V}$ were also found in other
configurations of n$_{V}<=$1/4 per 2-d.u.c..

\subsection{Lattice polarization dependence on the concentration of vacancy in LAO surface}
For the configurations of n$_{V}<$1/4 per 2-d.u.c. the carrier
density at the interface is less than 0.5 e/2-d.u.c.. The polar
electrical field in LAO overlayer is screened partly and then the
potential gradient in LAO overlayer still remains. In
Fig.~\ref{figDOS:1} (a) one can see that the valence and
conduction bands shift towards higher energy layer-by-layer from
the interface to the surface. The polar distortions in LAO induced
by residual polar electrical field are shown in
Fig.~\ref{figDOS:1} (c). Summarizing for configurations of
n$_{V}<=$1/4  we found that with increasing the concentration of
oxygen vacancy from (4$\times$4) to (2$\times$2) the density of
carriers at the interface increases till 0.5e/2-d.u.c and the
potential gradient and the polar distortions reduce till vanish.

For the configurations of n$_{V}<$1/4, the thickness of LAO is
another factor which affects the density of carriers at the
interface. For a given concentration of oxygen vacancies, with
increasing the thickness of LAO the VBM of LAO shifts upwards
until it touches CBM of STO. This critical thickness of LAO,
denoted as t$_{c}$, is dependent on the concentration of vacancies
in the surface n$_{V}$. Once the thickness of LAO overlayer
overruns t$_{c}$ the VBM of LAO does not shift up anymore,
instead, more charge transfers from valence band of LAO surface to
the interface. According to the electrostatics we give an
expression of carrier density n$_{c}$ as a function of n$_{V}$ and
t (thickness of LAO overlayer)
\begin{eqnarray}
  \label{eq:1}
  n_{c}= \left\{ \begin{array} {r@{\quad \quad}l}
    \frac{1}{2} - \frac{A}{t-t_{0}} & t>t_{c} \\ \\   2n_{v} & t\leq\ t_{c}
    \end{array} \right. \\
  t_{c}= \frac{A}{1/2-2n_{V}}+t_{0} ,
\end{eqnarray}
where A=1.97, t$_{0}$=0.053 obtained  from our previous DFT study.
The scale of t and t$_{0}$ is unit-cell layer. For the
configuration of (3$\times$3) with one vacancy in the surface the
critical thickness equals 7 u.c in our DFT calculation and 7.1
u.c. from above formula.

\begin{figure}[htbp]
    \centering
    \includegraphics[width=0.45\textwidth]{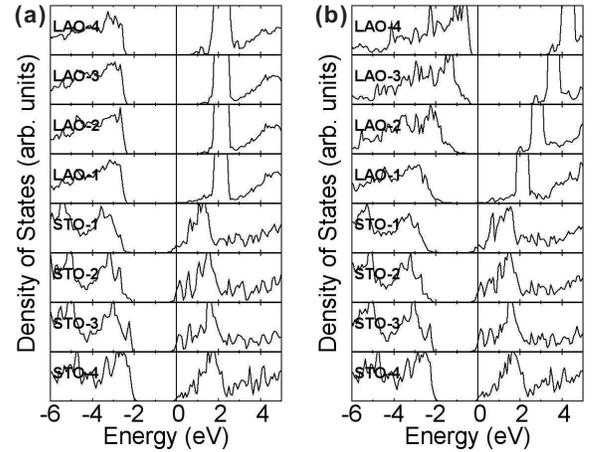}
    \caption{Layer-projected DOS for the (2$\times$2)
supercell (a) with one oxygen vacancy in LAO surface, and (b) with
one oxygen vacancy in STO-1 layer.} \label{figDOS:2}
\end{figure}

Figure~\ref{figDOS:2} (a) presents the layer-projected DOSs of
(2$\times$2) supercell with one vacancy in LAO surface. The state
of vacancy in LAO surface lies a little bit below the CBM of LAO
and is localized in the surface layer. Calculated density of
carriers at the interface is equal to 0.5 e/2-d.u.c., which is
exactly enough to completely screen the polar electric field in
LAO. One can see this point from straight-aligned VBMs and CBMs of
LAO layers in Fig.~\ref{figDOS:2} (a). Similar to what is shown in
Fig.~\ref{Struc} (b),  oxygen vacancy in the surface induces
strong tilt distortion which even extend into a few layers of STO,
but average polar displacement in each LAO layer is zero. It means
no net electrostatic field in LAO. For comparison
Fig.~\ref{figDOS:2} (b) presents the layer-projected DOSs of
(2$\times$2) supercell with one vacancy in STO substrate. Oxygen
vacancy in STO produces a shallow energy level in the CBM of STO
and generates two electron carriers. However, strong potential
gradient appears in LAO overlayer because no charge transfers from
the LAO surface to the interface.  We can conclude that carriers
generated by oxygen vacancies in STO hardly have any effect on
screening the polar electric field in LAO overlayer.

\section{Distribution of oxygen vacancies of high concentration}

\subsection{Binding energies}
\begin{figure}[htbp]
    \centering
    \includegraphics[width=0.45\textwidth]{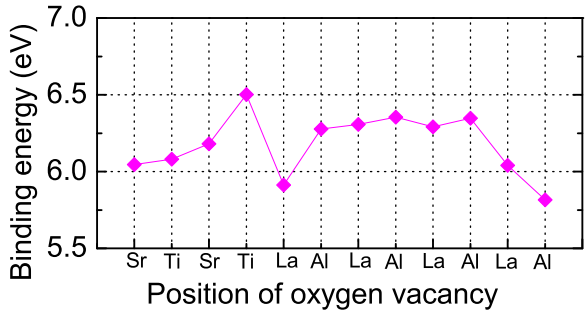}
    \caption{(Color online) Binding energies of  one oxygen
        atom at varying locations in (2$\times$2) supercell
        consisting of 4 u.c. layers of LAO under the condition
         of one oxygen vacancy fixed in LAO surface.}
        \label{BindingE2}
\end{figure}
We modeled the configurations of n$_{V}>$1/4 per 2-d.u.c. by
(2$\times$2) supercell with two oxygen vacancies. As we pointed
out above that when n$_{V}<=$1/4 per 2-d.u.c. oxygen vacancy
favors to lie in LAO surface. Considering this point in the
calculations of (2$\times$2) supercell with two oxygen vacancies
one is always fixed in LAO surface and the others varies from LAO
surface to STO subxtrate. Fig.~\ref{BindingE2} illustrates binding
energies of the oxygen atom corresponds to the second vacancy.
Although oxygen atom in surface still has the lowest binding
energy, in contrast to the cases of n$_{V}<=$1/4 per 2-d.u.c.  the
energy differences with other configurations decrease largely. The
energy difference between the configurations of the second vacancy
in LAO surface and in STO substrate is about 0.2 eV, which is far
less than that for the first vacancy. This implies that second
vacancy has large probability to appear in STO. Following
electronic structure calculations show that for the configurations
of n$_{V}>$1/4 per 2-d.u.c. the second vacancy does not contribute
any energy gain by screening the electrostatic field in LAO
because the first vacancy already completely screens it.


\begin{figure}[htbp]
    \centering
    \includegraphics[width=0.45\textwidth]{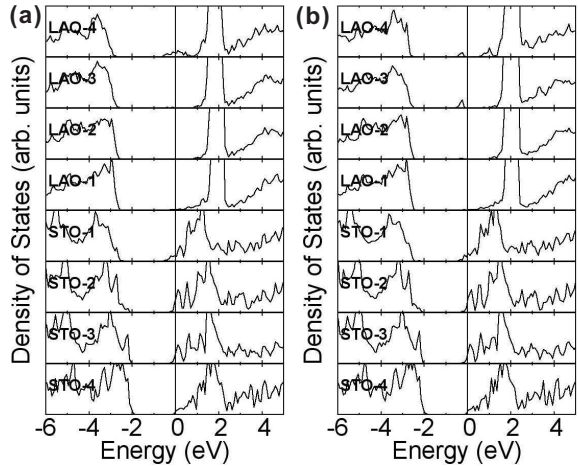}
    \caption{Layer-projected DOSs for (2$\times$2)
        supercells (a) with  two oxygen vacancies in LAO surface, and (b)
        with one vacancy in LAO surface and one in LAO-3 layer.}
\label{figDOS:3}
\end{figure}

\subsection{Electronic structure of  oxygen vacancies of high concentration}

Figure~\ref{figDOS:3} (a) presents the layer-projected DOS of
(2$\times$2) supercell with two oxygen vacancies in surface
AlO$_{2}$ layer.  The state of vacancies is widened due to the
interaction of two vacancies, and only half is occupied. Two
electrons in this state transfer to the interface. Another typical
configuration is shown in Fig.~\ref{figDOS:3}(b), in which one
vacancy lies in surface AlO$_{2}$ layer  and the other in the
 (LAO-3)-AlO$_{2}$ layer. The state of oxygen
vacancy in LAO-3 layer extends into the surface layer and is fully
occupied, while the state of oxygen vacancy in the surface layer
is empty. Our calculated amount of carriers at the interface is
still 2. In these two configurations the former has a metallic
surface, while the latter has an insulating surface. Both dope two
electron carrier at the interface, although two oxygen vacancies
exist in LAO. Several other configurations with two oxygen
vacancies in LAO also show the same carrier density 0.5e/2-d.u.c
at the interface. This indicates that oxygen vacancies in LAO can
contribute at most 0.5 electron carrier per 2-d.u.c. at the
interface, and extra vacancies apart from 1/4 per 2-d.u.c. have no
contribution to the conductivity at the interface. It can be
understood from the sight of polar electric field in LAO. Charge
transfer of 0.5 e/2-d.u.c from the LAO surface to the interface
completely compensate the polar field in LAO, lowering the total
energy. While Charge transfer of more than 0.5 e/2-d.u.c would set
up an inverse electric field in LAO, which could raise the total
energy.

\begin{figure}
    \centering
    \includegraphics[width=0.45\textwidth]{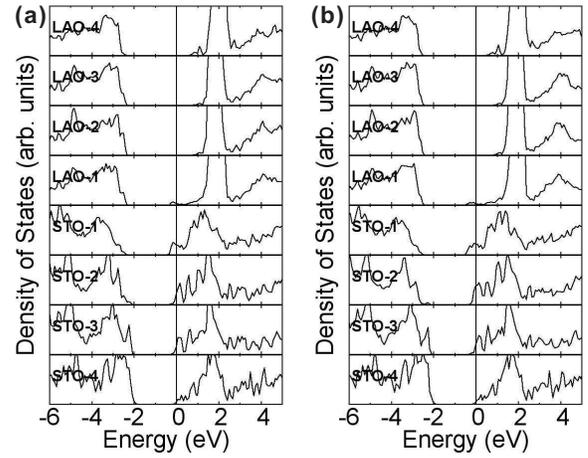}
    \caption{Layer-projected DOSs for the (2$\times$2)
supercells (a) with one oxygen vacancy in LAO surface and one in
STO, and (b) with one oxygen vacancy in LAO surface and two in
STO.} \label{figDOS:4}
\end{figure}

Since a great amount of oxygen vacancies were found in STO
substrate in the samples produced under lower oxygen pressure, the
contribution of vacancies in STO to density and distribution of
the carriers should be investigated.  Fig.~\ref{figDOS:4}(a) shows
the Layer-projected DOSs for the (2$\times$2) supercell with one
oxygen vacancy in LAO surface and one in STO. The carrier density
is 1.0 e/2-d.u.c. in our calculation. With one more vacancy in
STO, as shown in Fig.~\ref{figDOS:4}(b), the carrier density
increases  0.5 e/2-d.u.c. more. In contrast with the upper limit
contribution of the vacancies in LAO every vacancy in STO
generates two electron carriers in STO. Considering the vacancies
distributing uniformly in whole STO substrate under low oxygen
pressure in experiments, a 3-dimensional  distribution of carriers
can be formed. In respect of band structure no potential gradient
appears in LAO, while valence and conduction bands of STO have
evident bending at the interface.

\section{Discussion}
\subsection{Band diagrams}

\begin{figure}[htbp]
    \centering
    \includegraphics[width=0.45\textwidth]{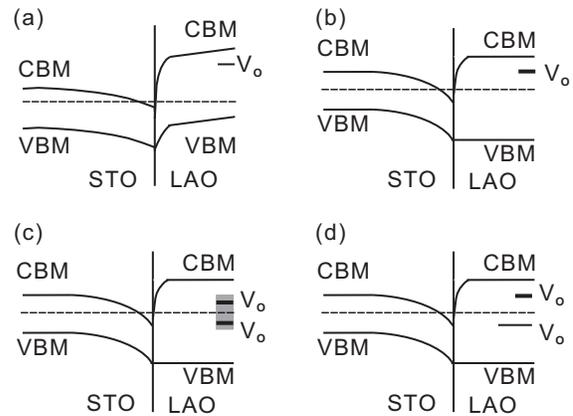}
    \caption{Band diagram of the interface of LAO overlayer on STO  with oxygen vacancies
    in LAO surface for  (a) n$_{V}<$ 1/4, (b) n$_{V}=$ 1/4, (c) n$_{V}>$ 1/4,
    (d) n$_{V}>$ 1/4 but vacancies in both surface layer
    and the layer below surface. } \label{figband}
\end{figure}

According to above calculated electronic structures we plot
schematic band diagrams of the interface. Since oxygen vacancies
in STO do not cause the band bending at the interface, only the
band structure with various concentrations of oxygen vacancies in
LAO are plotted.  Fig.~\ref{figband} presents the band structures
of oxygen vacancies in or near LAO surface. At the interface in
STO side strong band bending happens in all configurations. The
reason is that net positive-charged LAO overlayer, produced by
electrons transfer from LAO surface to STO, generates an
attractive potential to electrons in STO, lowing the potential at
the interface in STO side. In Fig.~\ref{figband} (a) n$_{V}<$1/4
per 2-d.u.c. and n$_{c}<$ 0.5 e/2-d.u.c.. The polar electrical  in
LAO is partially screened and the residual field results in bands
sloping in LAO. In Fig.~\ref{figband} (b) n$_{V}=$1/4 per 2-d.u.c.
and n$_{c}=$ 0.5 e/2-d.u.c.. The polar field is completely
screened, leading to flat bands in LAO. In contrast with
n$_{V}<$1/4 per 2-d.u.c. in the case of n$_{V}=$1/4 per 2-d.u.c.
more charge transfers from the surface to the interface, leading
to larger band bending at the interface in STO side. In
Fig.~\ref{figband} (c) n$_{V}>$1/4 per 2-d.u.c. but still n$_{c}=$
0.5 e/2-d.u.c.. The band structure is the same as the case of
n$_{V}=$1/4 but the surface is metallic. In Fig.~\ref{figband} (d)
vacancies are in surface and the layer below surface and the
surface is insulating. In previous XPS experiment a flat valence
band was observed\cite{yoshimatsu08}. According to above analysis
one can easily realize that such a band structure is arisen from
the oxygen vacancies of n$_{V}>=$1/4 per 2-d.u.c. in LAO surface.

\subsection{Charge carriers distribution: 2-dimensional or 3-dimensional}
\begin{figure}[htbp]
    \centering
    \includegraphics[width=0.45\textwidth]{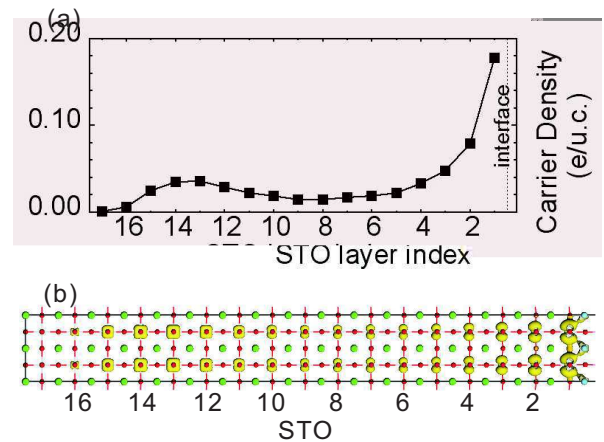}
    \caption{(a) Layer-resolved carrier density in STO
        for (LAO)$_{3}$/(STO)$_{17}$-(2$\times$1) with one oxygen vacancy in LAO surface.
        (b) Charge density plot of the induced carrier.} \label{Carrier}
\end{figure}

To clarify the distribution of the carriers induced by oxygen
vacancy we carried out the calculations of electronic structure
for (LAO)$_3$/(STO)$_{17}$-(2$\times$1) supercell by employing 17
STO layer. With one oxygen vacancy in LAO surface the integrated
carrier density in STO is 0.5 e/2-d.u.c.. While the carriers in
STO consist of two separate components, as illustrated in
Fig.~\ref{Carrier}(a). The interface component contains the main
part of carriers, and the extended part with a broad peak locates
near the 14th layer, i.e. about 5 nm, from the interface. The
potential profile in the STO side illustrated in
Fig.~\ref{figband} resembles the case of the inversion layer in
metal-oxide-semiconductor field-effect transistor and
semiconductor heterostructures\cite{ando82}. Within this
triangle-like potential, the carriers in the lowest bound state
which can be approximately described by the Airy function
accumulate at a few nm away from the interface\cite{ando82}.
However, the interface state is originated from different
mechanism. To understand the nature of two different components of
the carriers, we plotted the charge density of the induced carrier
in the STO side in Fig.~\ref{Carrier}(b). The interface component
consists mostly of the Ti $d_{xy}$ orbitals, but the extended
component has contributions from all the Ti $t_{2g}$ orbitals,
i.e., $d_{xy}$, $d_{yz}$, and $d_{zx}$ states. The character at
the interface stems from a strong compressive distortion of the
TiO$_{6}$ octahedron at the interface, which produces a strong
tetragonal field, lowering of the $d_{xy}$ state. Similar
character of carriers was found in the ideal interface of LAO/STO
without vacancy, and the mechanism was detailedly analyzed in our
previous work\cite{li09}.

From Fig.~\ref{Carrier} one can see that most of carriers
accumulate at the interface in four layers of STO although there
is extended state. Such a two-dimensional distribution of carriers
induced by oxygen vacancy in LAO surface is consistent with the
two-dimensional electron gas measured in LAO/STO samples produced
under higher oxygen pressure\cite{ohtomo04,basletic08}. While
under low oxygen pressure oxygen vacancies can be formed uniformly
in STO besides LAO surface. As well known, oxygen vacancies in STO
generate a shallow level in conduction band and Fermi level lies
in the CBM of STO. Considering band-bending at the interface the
carriers first fill in the potential valley at the interface and
then in STO inside. This implies that large amount of oxygen
vacancies in LAO/STO system would generate a three dimensional
distribution of carriers in STO but with higher carrier density
near the interface. Above picture about carrier distribution was
measured in previous experiment by using conducting atomic forcer
microscopy (AFM), in which the carrier density was found to decays
exponential near the interface and extends a few $\mu$m in
STO\cite{basletic08}.

\subsection{Localized surface conductivity}
Another finding in our calculations which we like to emphasize is
the surface conductivity in the system of LAO/STO. As shown in
Fig.~\ref{figDOS:3}(a), oxygen vacancies of more than 1/4 per
2-d.u.c. generated a metallic surface. In contrast to nearly
uniform distribution of the carriers at the interface, the
carriers in the surface is relatively localized. In the case of
Fig.~\ref{figDOS:3}(a), the states of two close vacancies overlap.
A part of electrons in the states transfer to the interface and
the left form localized surface carriers. This localized surface
carriers may be responsible for the few-nanometer-size conducting
islands in the surface of LAO/STO found in recent
experiment\cite{cen08}.

\section{Summary}
We investigated the n-type LAO overlayer on STO(001) substrate
with oxygen vacancies by using DFT calculation. We found that
oxygen vacancies favor to appear first in LAO surface, which
generate a two-dimensional carriers at the interface. The density
of carriers induced by vacancies in LAO surface has an upper limit
0.5 e/2-d.u.c.. For the concentration of vacancies in LAO surface
less than 1/4 per 2-d.u.c., the density of induced carriers at the
interface is less than 0.5 e/2-d.u.c. and the energy bands in LAO
slope up from interface to surface. While for that of equal to or
more than 1/4 per 2-d.u.c. the density of induced carriers is
equal to 0.5 e/2-d.u.c. and no slope of the energy bands occurs in
LAO. We found that for the case of the concentration of oxygen
vacancies more than 1/4 per 2-d.u.c. the surface presents
metallicity. We also investigated the role of oxygen vacancy in
STO. We found that every oxygen vacancy in STO generates two
electron carrier, but this carrier charge has no effect on
screening the polar electric field in LAO. We predict that when a
large amount of oxygen vacancies present in LAO/STO system the
carriers in STO show a three-dimensional distribution with higher
density at the interface.

$  $

\acknowledgements
This work was supported by BK21 and KOSEF
through the ARP (R17-2008-033-01000-0).  We also acknowledge the
support from KISTI under the Supercomputing Application Support
Program.

\end{document}